\documentclass[journal=ancac3,manuscript=article]{achemso}
\usepackage{amsmath}
\usepackage{amssymb}
\usepackage{color}
\usepackage{graphics}
\usepackage{epsfig}

\author{Juan L. Garcia-Pomar}

\affiliation{Instituto de Ciencia de Materiales de Arag\'{o}n and
Departamento de F\'{i}sica de la Materia Condensada,
CSIC-Universidad de Zaragoza, E-50009, Zaragoza, Spain}

\author{Alexey Yu. Nikitin}

\affiliation{Instituto de Ciencia de Materiales de Arag\'{o}n and Departamento de F\'{i}sica de la Materia Condensada,
CSIC-Universidad de Zaragoza, E-50009, Zaragoza, Spain}
\altaffiliation{Current address: IKERBASQUE, Basque Foundation for Science, 48011 Bilbao, Spain;
                CIC nanoGUNE Consolider, 20018 Donostia-San Sebasti\'{a}n, Spain}

\author{Luis Martin-Moreno}
\email{lmm@unizar.es}

\affiliation{Instituto de Ciencia de Materiales de Arag\'{o}n and Departamento de F\'{i}sica de la Materia Condensada,
CSIC-Universidad de Zaragoza, E-50009, Zaragoza, Spain}

\title{Scattering of Graphene plasmons by defects in the graphene sheet.}

\abbreviations{GSP, SPP, RPWEF, FEM, THz}

\begin{document}
\begin{abstract}
A theoretical study is presented on the scattering of graphene surface plasmons by defects in the graphene sheet they propagate in. These defects can be either natural (as domain boundaries, ripples and cracks, among others) or induced by an external gate. The scattering is shown to be governed by an integral equation, derived from a plane wave expansion of the fields, which in general must be solved numerically but it provides useful analytical results for small defects. Two main cases are considered: smooth variations of the graphene conductivity (characterized by a Gaussian conductivity profile) and sharp variations (represented by islands with different conductivity). In general, reflection largely dominates over radiation out of the graphene sheet. However, in the case of sharply defined conductivity islands there are some values of island size and frequency where the reflectance vanishes and, correspondingly, the radiation out of plane is the main scattering process.  For smooth defects, the reflectance spectra present a single maximum at the condition $k_p a \approx \sqrt{2}$, where $k_p$ is the GSP wavevector and $a$ the spatial width of the defect. In contrast, the reflectance spectra of sharp defects present periodic oscillations with period $k_p' a$, where $k_p'$ is the GSP wavelength inside the defect. Finally, the case of cracks (gaps in the graphene conductivity) is considered, showing that the reflectance is practically unity for gap widths larger than one tenth of the GSP wavelength.
\end{abstract}


\section{Introduction}

In the last few years, it has become evident that graphene not only displays remarkable electronic properties but can also play a significant role in photonics \cite{FerrariNP,Bao}. One aspect that has recently attracted much interest is that doped graphene supports bound electromagnetic modes, known as surface plasmons (GSPs)\cite{NovoselovNP}, that have the appealing characteristics of being both confined in a length scale much smaller than the free space wavelength\cite{Shung,Hanson,Jablan,Vakil2011}, and potentially controllable using external gates. Very recently, the existence of highly confined GSPs has received experimental confirmation. \cite{ju2011graphene,BasovNat12,KoppensNat12}.
Several aspects of GSP have already been studied theoretically, such as the efficient and directional coupling with nano emitters (and the associated enhanced spontaneous emission rate) \cite{Nikitin2011a,Koppens,EfimovPRB11,StauberPRB11,HansonAP11}, enhanced absorption and resonance diffraction\cite{deAbajoPRL12,NikitinPRB12,dielectrPRB12,EnhancedAPL12,Vasilevskiy12, Manjavacas2012},  metamaterials and antenna applications\cite{APLrings,JulianPerAPL12, Fang2012} and their wave guiding capabilities in ribbons\cite{Brey,Silvestrov,Popov,Nikitin2011,Andersen,Christensen} and edges \cite{Nikitin2011,Wang}.

However, very little is known about how GSPs behave when they encounter defects in the graphene sheet they propagate in. These defects can occur both (i) naturally as, for instance,
kinks appearing due to fabrication process \cite{Nagase}, domain borders in graphene growth by CVD\cite{ahmad}, the presence of multilayer islands\cite{Nagase}, cracks\cite{Kim,Zande} and different domains in CVD graphene\cite{ahmad} or (ii) be created externally, for example as the changes of conductivity in
gate-induced p-n\cite{lemme} or p-n-p junctions \cite{Gorbachev, liu}.

In this article, we present a theoretical study of GSP scattering by one-dimensional conductivity inhomogeneities. Calculations are conducted with an original method based on the Rayleigh expansion, which has the advantage of providing analytical expressions in some limiting cases.

\section{Model}
\begin{figure*}
  \includegraphics[width=16cm]{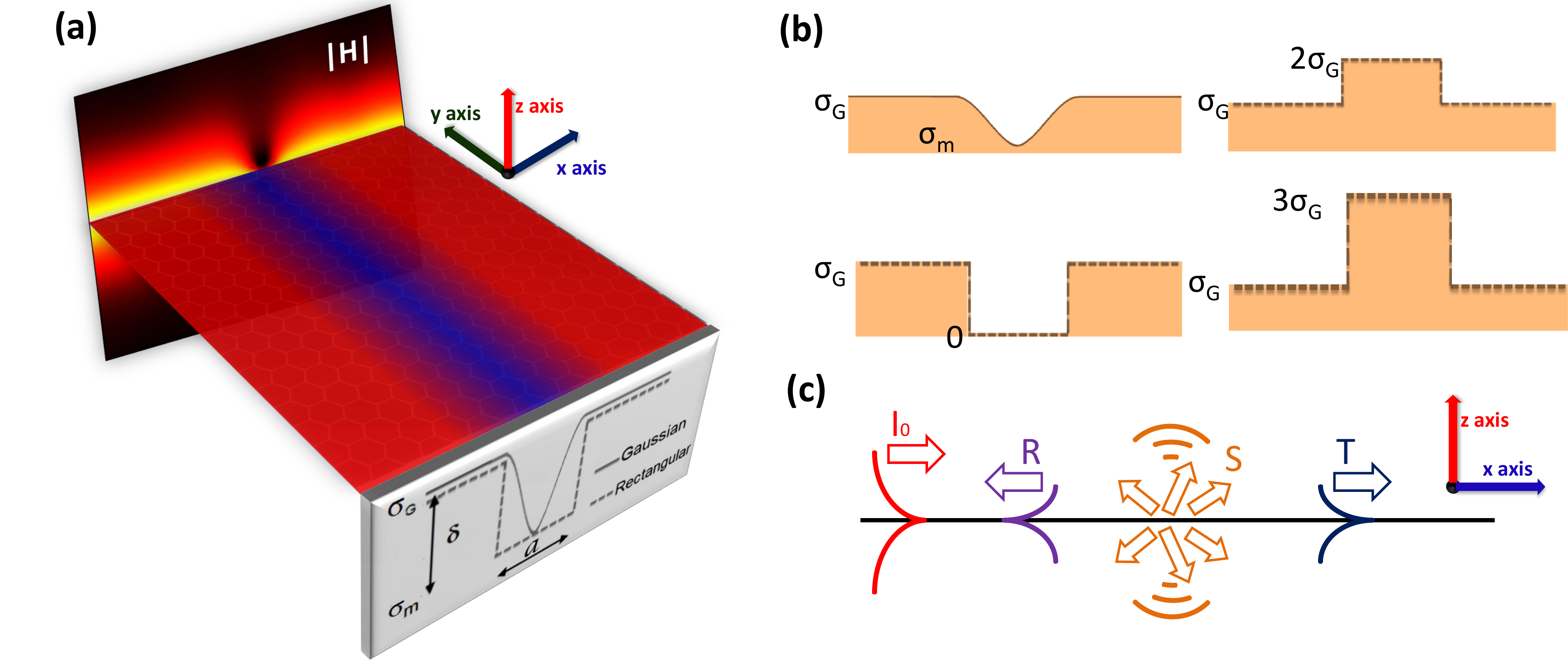}\\
  \caption{(Color Online) (a) Schematic geometry of free-standing graphene with a defect zone (blue color) that interacts with a GSP. The figure shows the conductivity profile (bottom) and a representative case of the computed the total (incident + scattered) magnetic field modulus $|H|$ (top). (b) Diagram for the different types of conductivity defects analyzed. (c) Schematics of the scattering processes occurring when a conductivity defect is present in a graphene sheet.}\label{fig1}
\end{figure*}

We consider a free-standing graphene monolayer, placed at $z=0$, with a spatial inhomogeneity in the two-dimensional conductivity $\sigma$. Actually, the presence of a substrate may be indispensable for applications, but it does not change any of the fundamental scattering properties of GSPs (affecting mainly the mobility of the charge carriers), which is why in this paper we concentrate on the simplest structure.  We will analyze one-dimensional (1D) inhomogeneities, with translational symmetry in the in-plane direction perpendicular to GSP incidence (the  $y$-direction), so $\sigma=\sigma(x)$. The geometry of the system is schematically shown in Fig. \ref{fig1}.

Away from the defect the (frequency-dependent) conductivity of graphene is $\sigma_G$. Defects have a characteristic width $a$, and a conductivity $\sigma_m$ at the defect centre ($x=0$) which can, alternatively, be described by the relative change in conductivity
$\delta=(\sigma_m-\sigma_G) /\sigma_G$.

In the scattering geometry, a monochromatic GSP (time dependency $e^{-i\omega t}$), propagating along the $Ox$ axis  (from the region $x<0$), impinges the defect, which induces some reflection back into the SPP channel, as well as some radiation out of the graphene sheet. Notice that, due to the symmetry of the problem, all scattered waves have the same polarization as the GSP (transverse magnetic). The scattering amplitudes can be computed by using numerical solvers of Maxwell equations.  Here, we present an alternative method, based on the Rayleigh plane-wave expansion (RPWE)\cite{Maradudin,Nikitin2007}, which in general must also be solved numerically but presents the advantage of providing analytical expressions for the scattering coefficients in some limiting cases. We leave all derivations for the Supporting Info and present here the main equations. Within the RPWE method, the electromagnetic field is written (all other components can be readily obtained from Maxwell equations) as:
\begin{equation}\label{eq1}
    E_x(x,z=0)=e^{iq_pgx}+\int^{\infty}_{-\infty}G(q) \, B(q)\, e^{iqgx} \, dq
\end{equation}
where $q$ and $q_p$ are $x$-components of the wavevectors of a plane wave and GSP respectively, normalized to the wavevector in vacuum $g=2\pi/\lambda$. $B(q)$ are scattering amplitudes, which satisfy the integral
equation:
\begin{equation}\label{eq2}
    B(q) =-\Delta\alpha(q-q_p) - \int^{\infty}_{-\infty}\Delta\alpha(q-q') \, G(q') \, B(q')dq'
\end{equation}
Here $\Delta\alpha(q)$ is the scattering potential related to the Fourier transform of the inhomogeneity of the dimensionless conductivity $\Delta \alpha(x)= (2 \pi/c) (\sigma(x)-\sigma_G)$ and
$G(q)=q_z/(1+q_z\alpha_G)$, where $c$ is the speed of light and $q_z=\sqrt{1-q^2}$.

The GSP reflectance and transmittance ($R$ and $T$ , respectively), as well as the fraction of energy flux scattered out of plane $S$ can be obtained from the  amplitudes $B(q)$ (see Supporting info). For instance,
$R =\left|2 \pi B(-q_p) /(q_p \alpha_G^3)\right|^2$.

It must be pointed out that this model does, in principle, account for losses in the graphene sheet. However, for the frequencies and defects that will be analyzed in this paper, for which $a$ is much smaller than the GSP absorption length $L_p$, the inclusion of losses leaves the scattering coefficients virtually unaltered. In a real situation in order to obtain reflexion/transmission coefficients, one has to normalize the GSP amplitudes by the decay factor $e^{-d/L_p}$, where $d$ is the distance that GSP runs from the launching point to the destination where the amplitude of the scattered GSP is measured. We have checked (running simulations with both the RPWE method and a commercial finite-elements code\cite{Comsol}) that once this procedure is applied for the lossy case, the scattering coefficients virtually coincide with those obtained when losses are neglected.  Therefore, and in order to concentrate on the scattering coefficients intrinsically due to the defect, all calculations presented in this paper have been obtained setting $\mathrm{Re}[\sigma_G]=0$. In this way, current conservation implies $R+T+S=1$. Throughout the paper the conductivity is taken from the RPA expression\cite{Wunsch,Hwang,Falkovsky} and, for definiteness, we consider that the chemical potential is $\mu=0.2$eV. As we will show, this choice is not essential, as most results only depend on $\mu$ through the value of the GSP wavevector.

Of course, for the scattering coefficients associated to any particular defect will depend on its conductivity profile. Here we do not attempt to computing this profile; instead, we will assume some basic spatial dependences for the conductivity and compute how they scatter GSPs.
We analyze two differentiated main cases: (i) smooth variations in the conductivity, described by a Gaussian profile  $  \sigma(x)=\sigma_G\{1+ \delta \, \exp(-4x^{2}/a^2)\}$, where $a$ is the full spatial width at $1/e$ relative conductivity change, and (ii) abrupt ones, represented by a step defect  $\sigma(x)=\sigma_G\{1+ \delta \, \Theta(a/2-|x|)\}$, where $\Theta(x)$ is the Heaviside step function.

\section{Smooth defects}

Let us advance that for, all smooth defects considered, our calculations show that the scattering out-of-plane is extremely small ($S/R \lesssim10^{-4}-10^{-2}$). This point will be discussed later on; now it allows us to
focus on the reflectance, from where the transmittance can be obtained as $T\approx 1-R$.

We first consider a Gaussian profile variation in the conductivity with a small $\delta$. This profile provides a good approximation to the realistic variations on graphene conductivity arising from atomic steps in the substrate, when graphene is grown on SiC\cite{Nagase}. In this case the characteristic width of the Gaussian profile is $a\sim 20-50nm$, and the conductivity relative change $\delta$ ranges between $-0.01$ and $-0.5$.

\begin{figure}
  \includegraphics[width=16cm]{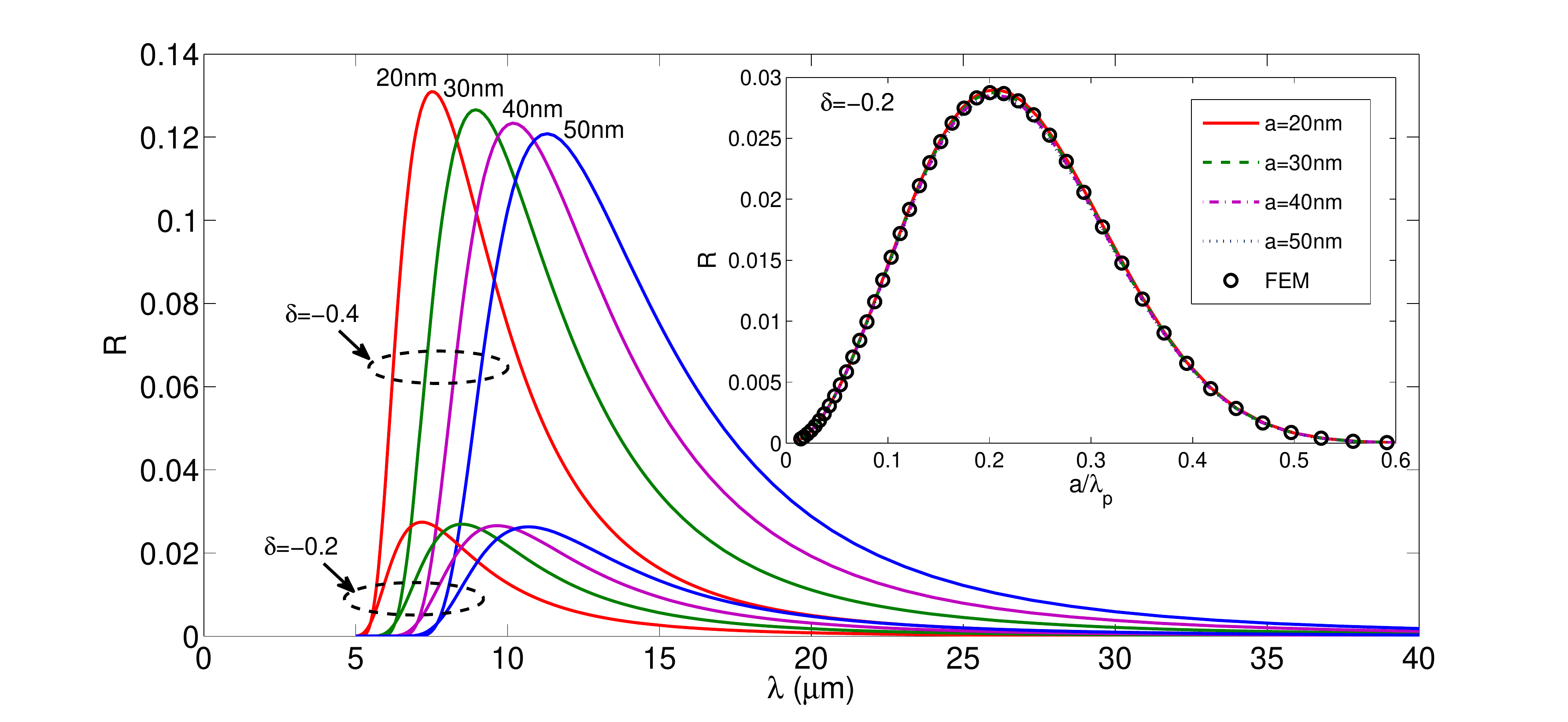}\\
  \caption{Reflectance spectra $R(\lambda)$ for a GSP impinging a shallow Gaussian conductivity defect, for different defect widths and two values of $\delta$. Inset: R as a function of $a/\lambda_p$ for $\delta=-0.2$ and different widths, showing that the reflectance scales with $a/\lambda_p$. The results obtained with the finite element method (open circles) for $a=30nm$ and $\delta=-0.2$ are also shown, in order to confirm the validity of the RPWE calculations.}\label{fig2}
\end{figure}

Fig. \ref{fig2} renders the reflectance spectra $R$ for defects with different widths and two values of the relative change in conductivity. We observe that $R$ has a maximum for each value of $\delta$ and $a$. This maximum arises as a compromise between two distinct asymptotic dependencies. For small GSP wavelengths, $\lambda_{p}$, the GSP follows adiabatically the variation of the conductivity and virtually no reflection is generated. Conversely, in the long-wavelength region, the defect width is very small in relative terms ($a\ll\lambda_{p}$) and so is the reflectance, which decreases with $\lambda$ due to the decrease in $a/\lambda_{p}$. In between these asymptotic decays there is maximum for $R$, related to the $q$-space Fourier image of the Gaussian conductivity profile.

\begin{figure}
  \includegraphics[width=10cm]{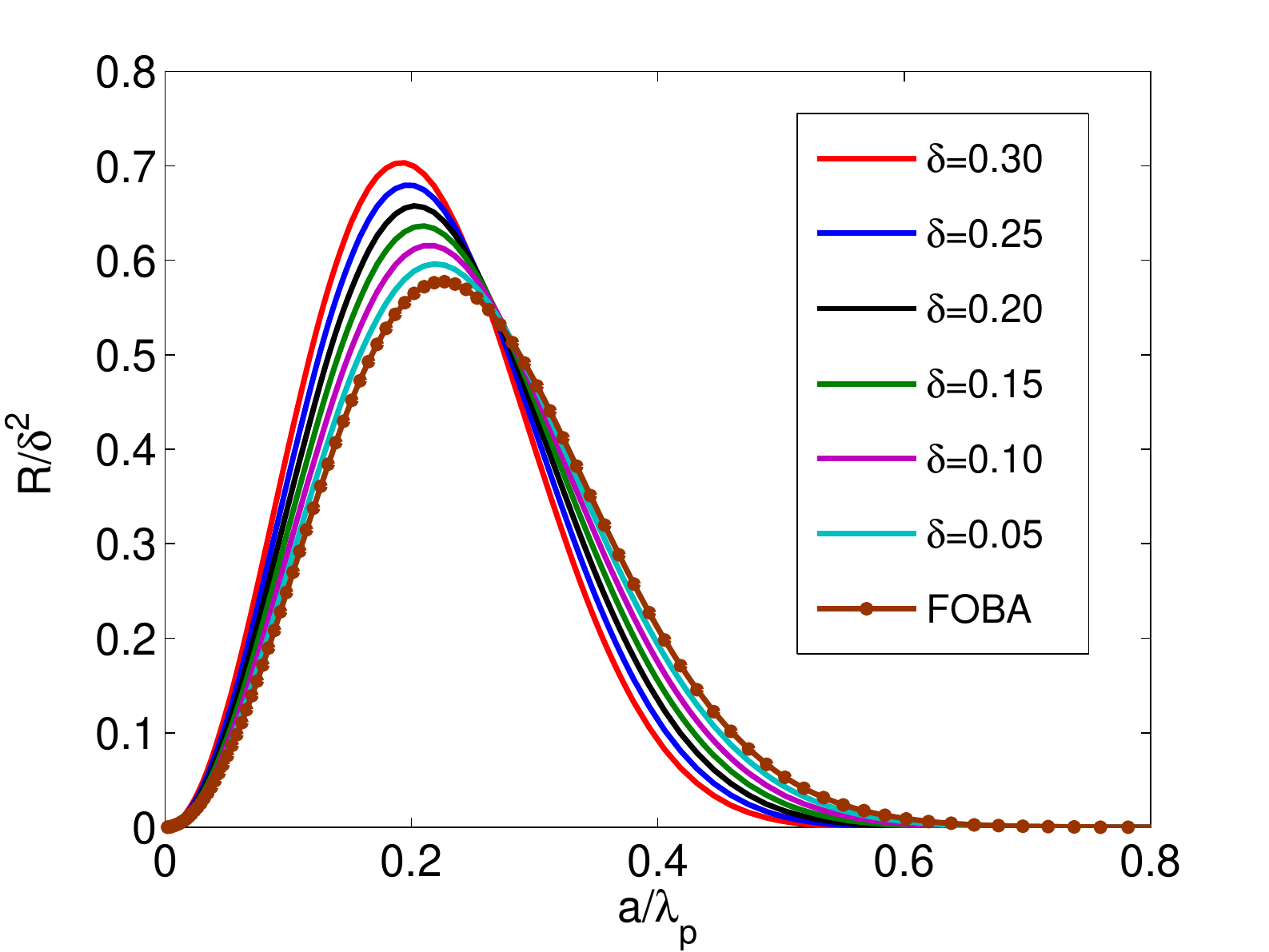}\\
  \caption{$R/\delta^2$ as function of $a/\lambda_p$, for for a Gaussian conductivity defect with $a=50nm$ and different values of $\delta$. We observe that, as $\delta$ decreases, the curves converge to the universal reflectance spectra predicted by the FOBA.}\label{fig3}
\end{figure}

In order to gain more insight into the behavior of the GSP reflectance, and obtain some quantitative estimations to support the exact calculation, we compute the plane-wave amplitudes $B(q)$ within the first-order Born approximation (FOBA). This approximation is valid for small variations of $\delta$ and
 corresponds to neglecting the integral term in the right-hand side of Eq.(2),
keeping only the linear term in $\Delta \alpha$.

Within FOBA the scattering amplitude reads $B^{FOBA}(q)=-\Delta\alpha(q-q_p)$. Using  $\alpha_G^{-2} = 1-q_p^2\approx -q_p^2$ we obtain (see details in Appendix)
 \begin{equation}\label{FOba3}
    R^{FOBA}=\frac{\pi}{4}  (k_p a)^2 \, e^{-\frac{1}{2} (k_p a)^2} \, \delta^2
 \end{equation}
 Notice that, within the FOBA, the quantity $R/\delta^2$ is a universal function of $a/\lambda_p$. It also predicts that the maximum in reflectance occurs when $a/\lambda_p = 1/( \sqrt{2} \pi) \approx 0.22$ (independent of $\delta$), with a maximum reflectance $R^{FOBA}_{max}=(\pi/2e) \delta^2 \approx 0.58 \delta^2$.
The validity of the scaling of the reflctance with $a/\lambda_p$ predicted by the FOBA is shown in the inset to Fig. \ref{fig2}, for different defect widths and $\delta=-0.2$.  Additionally, Fig. \ref{fig3}. renders the scaling with $\delta$ computed for a narrow defect, together with the prediction by the FOBA. The FOBA captures very accurately the spectral position of the maximum reflectance, even for moderate variations in conductivity, and gives a good approximation to the full reflectance spectra.
The FOBA also provides insight into the relative strength of reflectance and radiation channels. The scattering strength depends both on (i) the density of final states (which is much larger for GSPs than for radiation channels) and (ii) a matrix element, given by the Fourier component of the conductivity variation evaluated at the wavevector difference $\Delta k$ between the GSP one and that of the final state (i.e., $\Delta k = - 2 k_p$ for reflectance and $\Delta k \approx - k_p$ for radiation processes). In the case of smooth defects, the FOBA shows that the density-of-states factor dominates over the "matrix element" one (see Appendix). Actually, at the reflectance maximum, the FOBA predicts $S/R = 0.5 e^{3/4} |\alpha_G|^{2}$, which is in the range  $ \sim 10^{-4}-10^{-2}$ for the values of $ \alpha_G$ relevant for GSP propagation ($\alpha_G  \sim 1-5 \alpha_0$, where $\alpha_0 \approx 1/137 $ is the fine structure constant). This preponderance of $R$ over $S$ holds even for larger defects strengths, where the FOBA is no longer strictly applicable.

It is also interesting to analyze the scattering by smooth defects with large $\delta$, as they can be produced by changing the carrier concentration in graphene (and thus the conductivity) with an external gate\cite{Gorbachev, liu}. The Gaussian shape in this case may simulate a  $n^--n-n^-$ (or $p^+-p-p^+$) junction. In this case, the defect width will depend on the geometrical details of the gate, but can be expected to be of order of $0.2-1$ $\mu m$, and $\delta$ can be considered a tunable parameter ranging from $-1$ to $0$.

\begin{figure}
  \includegraphics[width=13cm]{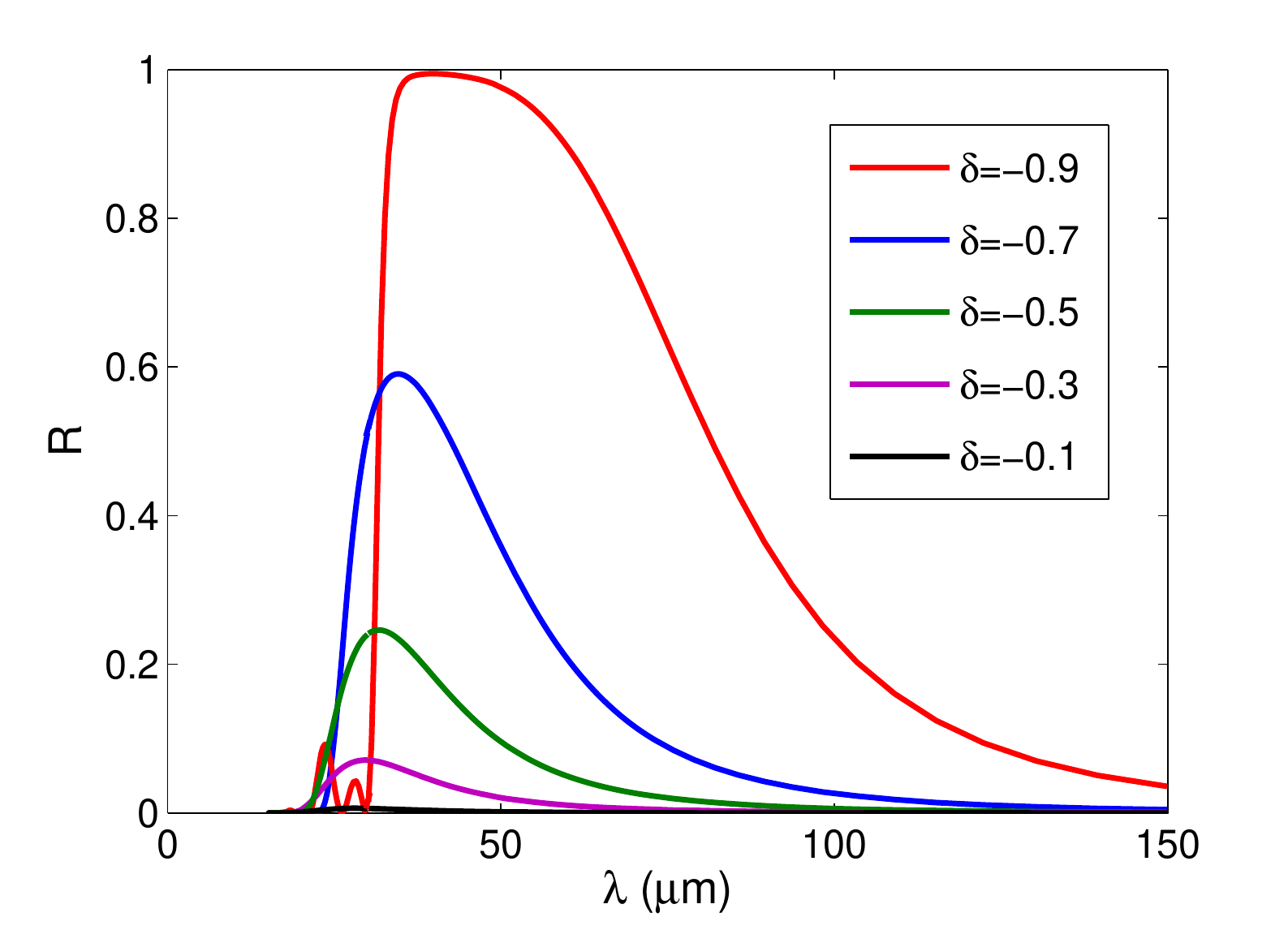}\\
  \caption{Reflectance spectra for a GSP impinging a Gaussian conductivity defect for different values of $\delta$. The defect width is $a= 400nm$.}\label{fig4}
\end{figure}

\begin{figure}[htb]
\includegraphics [width=13cm]{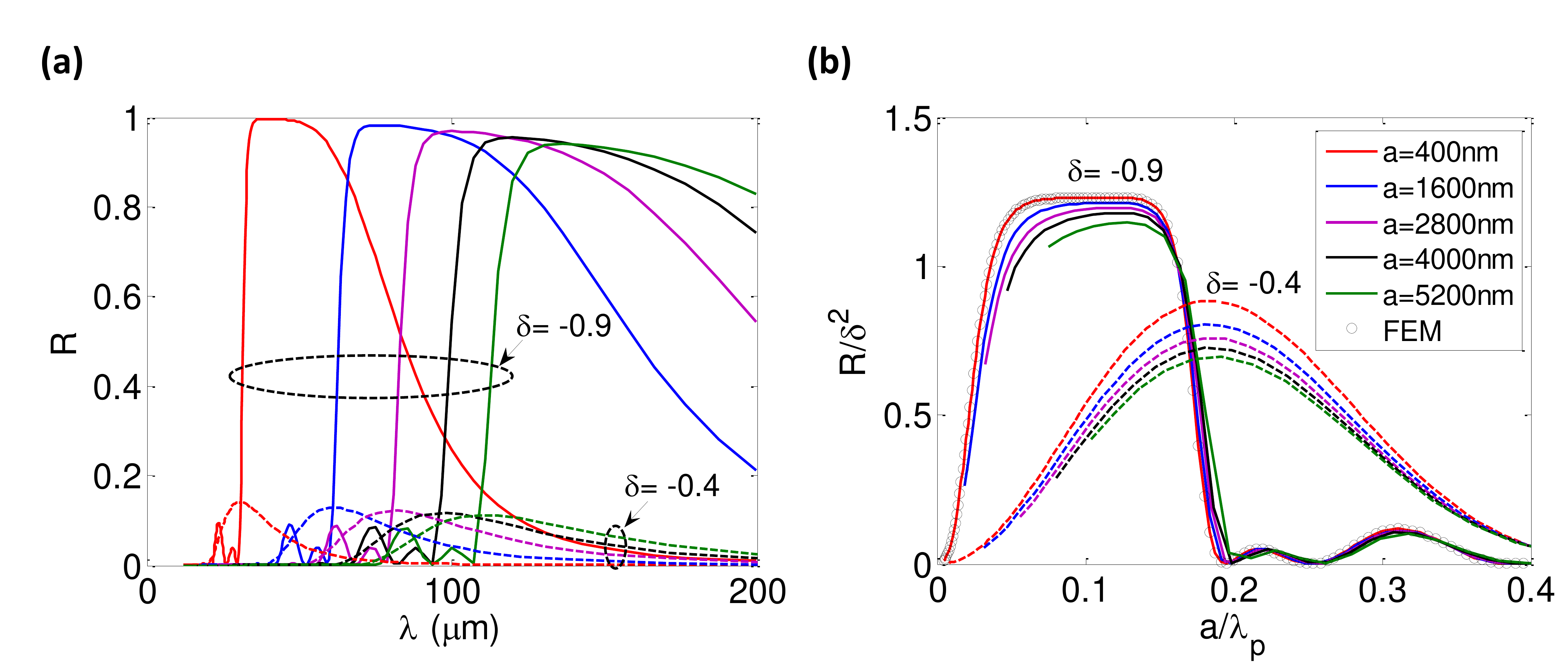}
    \caption{(a) Reflectance spectra for a GSP impinging a Gaussian conductivity defect, for different widths and for both $\delta=-0.9$ and $\delta=-0.4$ . (b) $R/\delta^2$ in function of $a/\lambda_p$ for the same parameters considered in (a), showing the approximate scaling of the reflectance spectra. The open circles render calculations performed within the FEM (in order to validate those performed with the RPWE method), for $a=400nm$ and $\delta=-0.9$.}
    \label{fig5}
\end{figure}

Fig. \ref{fig4} presents the reflectance spectra for different values of $\delta$, for the fixed defect width $a=400nm$. These results show that, for large relative changes of the conductivity, the maximum reflectance occurs for even smaller defect widths than those predicted by the FOBA and that, for large $|\delta|$, there are spectral regions where the reflectivity is high. This point is even more apparent in Fig. \ref{fig5}(a), which renders the reflectance spectra for $\delta=-0.9$ and different defect widths. Interestingly, the reflectance still satisfies approximately the scaling relation in $a/\lambda_p$ predicted by the FOBA, see Fig. \ref{fig5}(b), although for this large value of $|\delta|$ the FOBA is no longer a good approximation to the scattering amplitudes, which must be obtained by solving the full integral equation Eq.(2).

\begin{figure}[htb]
\includegraphics [width=16cm]{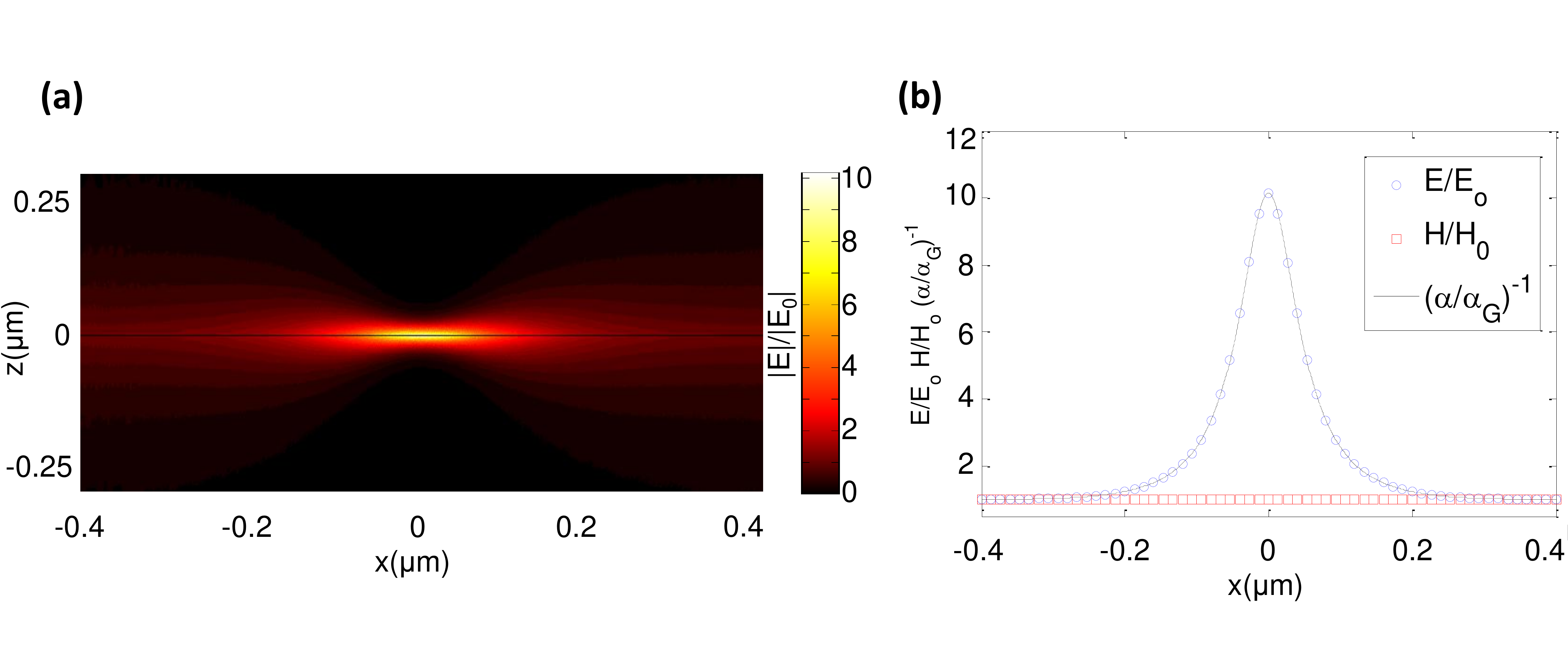}
  \caption{(a) Snapshot of the electric field norm along the GSP interacting with a Gaussian conductivity defect characterized by  $\delta=-0.9$, $a=200nm$ and $\lambda=200\mu m$. (b) Crosscut at z=0 of the previous snapshot. Open squares represent the normalized magnetic field norm $H/H_0$ (open squares), which is approximately unity along the GSP propagation, open blue circles render the normalized electric field norm $E/E_0$, showing the exact scaling with the inverse of normalized dimensionless conductivity $(\alpha/\alpha_G)^{-1}$ (continuous black line).}\label{fig6}
\end{figure}

It is remarkable that for the large relative change in conductivity considered in Fig. \ref{fig5}, the reflectance in the spectral region $a/\lambda_p > 0.2$ is small. Actually, there are values of $a/\lambda_p$ where the reflectance vanishes and, given the scattering out of plane is negligible, the transmittance is almost unity.
As the GSP extension in the direction perpendicular to the graphene sheet scales with the conductivity, the GSP is very strongly bound at the defect centre. Then, unit transmittance and conservation of energy imply that the electric field is strongly enhanced at the centre, as illustrated in Fig. \ref{fig6}(a). The scaling of the electric field amplitude can be obtained by assuming that, for adiabatic propagation of GSPs, the electrical current along the graphene sheet is constant, i.e. $|J(x)|=|\sigma(x)|\cdot |E_x(x)|=constant$, leading to $|E_x(x)|\propto 1/|\sigma(x)|$. As for the other EM components of the GSP field, we know that they satisfy $|E_x|\simeq|E_z|\simeq |q_pH|$. So, taking into account that locally $q_p(x)\simeq i/\alpha(x)$, we arrive at $|H(x)|\simeq |\alpha(x) E_x(x)|\propto constant$. The spatial dependence of the GSP field is rendered in Fig. \ref{fig6}(b), together with the conductivity profile, fully confirming the predicted scaling behavior.

\section{Abrupt defects}

\begin{figure}
  \includegraphics[width=13cm]{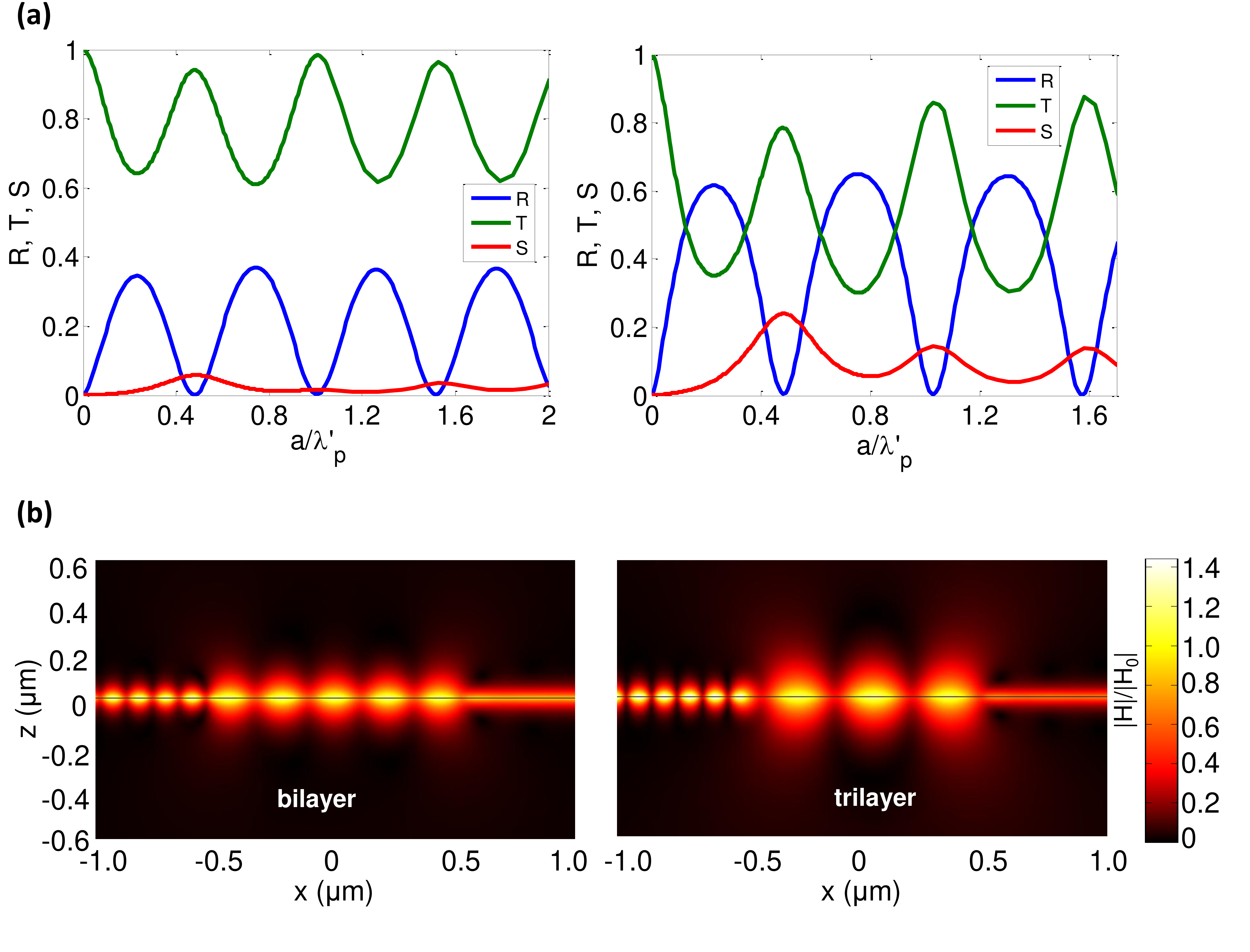}\\
  \caption{(a) Reflectance $R$, transmittance $T$ and fraction of energy radiated out of plane $S$ versus $a/\lambda'_p$ for a bilayer (left) and trilayer (right) for $\lambda=10\mu m$. (b) Snapshots of the magnetic field norm for a GSP propagating in a graphene monolayer and impinging onto either a bilayer (left) or a trilayer (right) strip. The strip width is $a=1\mu m$. }\label{fig7}
\end{figure}

One paradigmatic case of defect with abrupt change in conductivity is an island of multilayer graphene, placed on a  graphene monolayer. As for small number of layers the thickness of the multilayer is much smaller than the GSP extension along the normal to the sheet, this thickness can be neglected and the multilayer region can be approximated by a conductivity defect. Here we analyze the scattering by both bi- and tri- layer strips, and approximate their conductivities as twice ($\delta=1$) or three times ($\delta=2$) the conductivity of a monolayer, respectively.
The FOBA calculation for these rectangular-type defects gives
$ R^{FOBA}= \mathrm{sin}^2(k_p a) \, \delta^2$, predicting that  the reflectance spectra oscillates periodically when expressed as a function of $a/\lambda_p$. Within the FOBA, reflectance minima occur at $a= n\lambda_p/2$, $n=1,2,\ldots$ (which can be interpreted as the constructive interference in the backward direction between GSP partially reflected at the edges of the defect). The FOBA is not a good approximation for these defects where $\delta$ is not small, as it fails to take into account the modification in the GSP field inside the island due to the change in conductivity. However, we have found that the full calculations follow quite approximately the periodic behavior predicted by the FOBA, but as a function of $a/\lambda'_p$, where $\lambda'_p$ is the wavelength of the plasmon corresponding to the conductivity inside the defect, see Fig. \ref{fig7}(a). Notice also that, for these abrupt defects, the scattering out of plane also presents an oscillatory behavior. Its amplitude, although smaller than the one for the reflectance, is not negligible, and is peaked (as also does the transmittance) at the spectral positions where the reflectance is minimum.

As illustrated in Fig. \ref{fig7}(b), the strong reflection of GSP at the island boundaries result is the formation of standing waves in the island, with a number of nodes determined by both the island size and the GSP wavelength there. Notice that, as the conductivity inside the multilayer island is larger than in the monolayer, the GSP is less strongly bound to the graphene sheet. Also, Fig. \ref{fig7}(b) clearly shows that the out-of-plane radiation is generated at the island boundaries.

\begin{figure}
  \includegraphics[width=13cm]{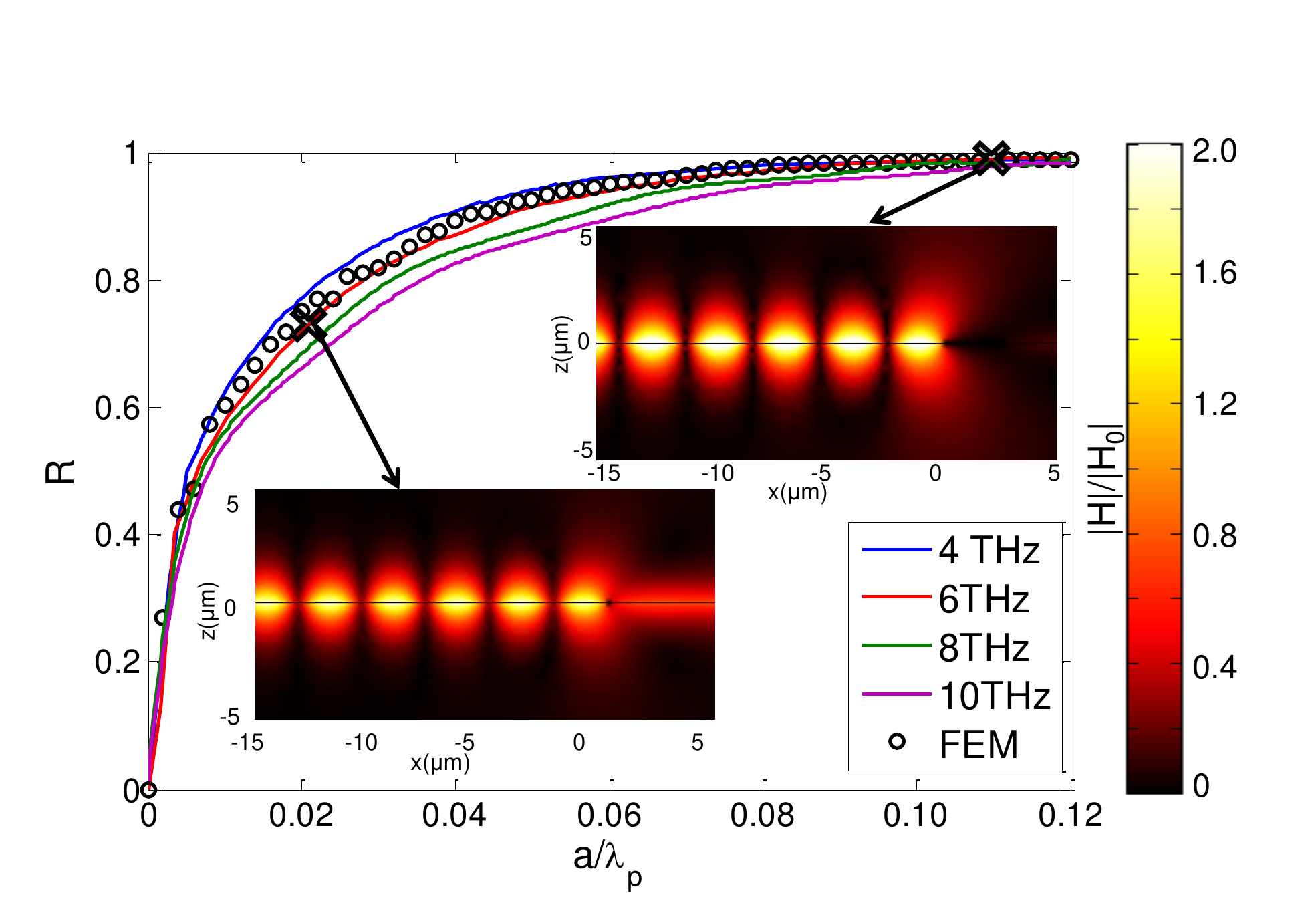}\\
  \caption{Reflectance $R$ for a GSP impinging onto a crack, as a function of crack width $a$, for different frequencies. FEM calculation for 6THz is shown on order to demonstrate the validity of the RPWE method. Inset: snapshots of the magnetic field norm, normalized to that of the incident magnetic field at the graphene sheet, for the cases $a/\lambda_p=0.02$ and $a/\lambda_p=0.11$.}\label{fig8}
\end{figure}

Finally, as another paradigmatic case of abrupt defect, we study the scattering of a GSP by a crack in the graphene layer. The exact spatial dependence of the conductivity near the graphene edge is a question still under debate, with the microscopic details of the graphene edge (whether has a zigzag or armchair configuration) possibly playing an important role\cite{owens,GdeAbajo2012}. Here we will simply assume that the conductivity vanishes within the gap region, in order to provide an estimation of the distances that GSP can tunnel through.
Fig. \ref{fig8} renders the dependence of the computed reflectance with gap width, for several frequencies. The results are presented as function of $a/\lambda_p$, showing that the reflectance approximately follows a scaling behavior. Still, the reflectivity is high already for small values of $a/\lambda_p$, demonstrating the extreme sensitivity of GSPs to the presence of cracks in the graphene sheet. The insets to Fig. \ref{fig8} show snapshots of the magnetic field, illustrating (i) the standing wave arising from the reflection of the GSP at the crack, (ii) the smallness of the radiation out of the graphene sheet (in all calculations, the fraction of GSP energy radiated out of plane is $S \approx 10^{-3}$), and (iii) that the GSP is fully reflected for widths $a>0.1\lambda_p$.

\section{Conclusion}

We have analyzed the scattering properties of GSP by defects in the local conductivity of the graphene sheet.
In the case of smooth spatial variations of the defect conductivity (which occur for instance when the defect is created by modification of the carrier concentration via a  top gate) we have found that, for a given relative change in the conductivity at the defect centre, the reflectance follows approximately a universal scaling in terms of $a/\lambda_p$. In all cases, the reflectance reaches its spectral maximum value when $a \approx 0.2  \lambda_{p}$.
When $a/\lambda_{p}$ is larger than that given from the previous condition the GSP propagation can be considered as adiabatic, and thus the GSP is mainly transmitted. When, additionally, the conductivity at the centre of the defect is small this leads to a strong electric field enhancement at the defect centre. We have also found that for these smooth defects, the scattering out of plane is always much smaller than the reflectance.

A different behavior is found for the scattering of GSP by multilayer islands, placed in a monolayer background. In this case, the scattering out of plane it is not negligible, although on average it is smaller than the reflectance. In fact, the reflectance spectra oscillates periodically as a function of quotient between the island width and the plasmon wavelength {\it inside the defect}, with both transmittance and scattering out of plane presenting maxima at the reflectance minima.

Finally we have found that conductivity gaps in the graphene sheet prevent very efficiently the GSP propagation, with the GSP being fully reflected for gap widths larger than $\sim 0.1 \lambda_p$.

\begin{acknowledgement}
This work has been partially funded by the Spanish Ministry of Science and Innovation under contracts MAT2011-28581-C02 and CSD2007-046-NanoLight.es.
\end{acknowledgement}

\appendix
\section{Appendix}
\setcounter{equation}{0}
\renewcommand{\theequation}{A{\arabic{equation}}}

- {\it Derivation of the integral equation governing scattering of GSPs by a defect.}

Let us assume that the normalized conductivity of graphene $\alpha=2\pi\sigma/c$ is the following function of the coordinate $x$: $\alpha(x)= \alpha_G+\Delta\alpha(x)$, where $\alpha_G$ is the background conductivity [$\alpha_G = \alpha(x\rightarrow\pm\infty)$] and $\Delta\alpha$ is a localized perturbation [$\Delta\alpha(x\rightarrow\pm\infty)=0$]. The perturbation can be represented by its Fourier expansion
\begin{equation}\label{sm1}
\begin{split}
\Delta\alpha (x) = \int dk \, \Delta\alpha(k) \, e^{ikx}, \hspace{0.2cm} \Delta\alpha(k) = \frac{1}{2\pi}\int dx \, \Delta\alpha(x) \, e^{-ikx}.
 \end{split}
\end{equation}

We consider a graphene surface plasmon (GSP) propagating along a free-standing graphene sheet located at $z=0$, from the region $x<0$ (we assume lossless graphene for the reasons mentioned in the manuscript). The GSP impinges the inhomogeneity $\Delta\alpha(x)$ localized around $x=0$ and is partially reflected, transmitted and scattered into a continuum of propagating and evanescent modes $|k>$. Then the longitudinal components of the electric fields above and below a monolayer can be exactly represented by the Rayleigh integral expansion [time dependency $\sim \exp(-i\omega t)$ is supposed everywhere]
\begin{equation}\label{sm2}
\begin{split}
&E_x(x,z)=E^+_x (x,z) = e^{ik_px} \, e^{ik_{zp}z} + \int_{-\infty}^{\infty} dk E^+(k)\,e^{ik_x}\, e^{ik_{z}z}, \hspace{0.2cm} z>0,\\
&E_x(x,z)=E^-_x (x,z) = e^{ik_px} \, e^{-ik_{zp}z} + \int_{-\infty}^{\infty} dk E^-(k)\,e^{ik_x}\,e^{-ik_{z}z}, \hspace{0.2cm} z<0.
  \end{split}
\end{equation}
with $k_p = g\sqrt{1-1/\alpha_G^2}$, $g=\omega/c= 2 \pi /\lambda$. The $z$-component of the wavevectors are given by $k_z=\sqrt{g^2-k^2}$ (with $\mathrm{Im}(k_z)>0$), and $k_{zp}=-g/\alpha_G$. The components of the electric field $\mathbf{E} = (E_x,0,E_z)$ and magnetic field $\mathbf{H}=(0,H_y,0)$ are connected through the Maxwell's equations.

The boundary conditions at $z=0$ are provided by (i) the continuity of the parallel component of the electric field and (ii) the jump of the parallel component of the magnetic field across the graphene layer due to the induced electric current
\begin{equation}\label{sm4a}
\begin{split}
&E_x^+ (x,0) = E_x^-(x,0),\\
&H_x^+ (x,0) - H_x^-(x,0) = 2\alpha(x) E_x^+ (x,0).
\end{split}
\end{equation}
The fields given by Eq.(A2) can be substituted into the boundary conditions Eq.(A3). Using Eq.(A1) and projecting the equations onto the basis set $e^{ik x}$ we arrive at the integral equation for the Fourier amplitude of the field $E^+(k)$:
\begin{equation}\label{sm4}
    B(q) =-\Delta\alpha(q-q_p) - \int\Delta\alpha(q-q')B(q')G(q')dq'
\end{equation}
Here we have introduced the normalized wavevector components $q=k/g$, $q_z=k_z/g$, so that $ \Delta\alpha(q) = g \Delta\alpha(k)$, and normalized the Fourier image of the field to the Green's function $G(q)=q_z/(1+\alpha_Gq_z)$:
\begin{equation}\label{sm5}
\begin{split}
E^+(k) = G(q) \, B(q).
\end{split}
\end{equation}

Eq.(A4) has been solved discretizing $q$ and replacing the infinite region of integration by increasingly larger finite limits, until convergency is achieved. A non-uniform discretization scheme in q-space is considered, in order to take into account the strong variations of the Green's function $G(q)$\cite{Maradudin,Nikitin2007}.

Once the Fourier image of the field is computed, we can obtain the GSP reflection, transmission, and out-of-plane scattering efficiency.
For this, we use the normalization Eq.(A5). Assuming that the function $B(q)$ does not have poles in the complex plane $q$,  the asymptotic (long-distance) behavior of the field at $x\rightarrow \pm\infty$ is given by the contribution from the poles of $G(q)$:
\begin{equation}\label{sm7}
\begin{split}
&E_x(x\rightarrow\infty,0)=(1+\tau)e^{ik_px}, \\
&E_x(x\rightarrow-\infty,0)= e^{ik_px}+\rho e^{-ik_px},
  \end{split}
\end{equation}
where
\begin{equation}\label{sm8}
\begin{split}
\tau = \frac{2\pi i}{\alpha_G^3q_p}B(q_p), \hspace{0.2cm} \rho = \frac{2\pi i}{\alpha_G^3q_p}B(-q_p).
  \end{split}
\end{equation}
From Eq.(A6) and Eq.(A7), the reflectance and transmittance are immediately obtained:
\begin{equation}\label{sm9}
\begin{split}
R = |\rho|^2=\left|\frac{2\pi i}{\alpha_G^3q_p}B(-q_p)\right|^2, \\
T = |1+\tau|^2=\left|1+\frac{2\pi i}{\alpha_G^3q_p}B(q_p)\right|^2.
  \end{split}
\end{equation}
The computation of the energy radiated in the form of the propagating waves is done by performing the saddle-point integration of the fields in the far-zone,  and integrating the normal component of the Poynting vector $\mathbf{P} = \frac{1}{2}\mathrm{Re}[\mathbf{E}\times \mathbf{H}^{\ast}]$ over a closed cylindrical surface. Then, normalizing it onto the flux of the incident SPP, $J_p = q_p|\alpha_G|^3/2g$, we obtain
\begin{equation}\label{sm10}
\begin{split}
S =\frac{4\pi}{q_p|\alpha_G|^3}\int_{|q|<1}\frac{dq}{q_z} |B(q)G(q)|^2.
  \end{split}
\end{equation}
which can also be written in a form that permits to define the angular dependence of the radiated flux:
\begin{equation}\label{sm11}
\begin{split}
S =\int_{-\pi/2}^{\pi/2}d\theta D(\theta),
  \end{split}
\end{equation}
where $D(\theta)$ is the differential cross section:
\begin{equation}\label{sm12}
    D(\theta)=\frac{4\pi}{q_p|\alpha_G|^3}|B(\cos\theta)G(\cos\theta)|^2,
\end{equation}
with the angle $\theta$ counted from the axis $Oz$.

As we have assumed that our system is lossless, the scattering coefficients constitute the current conservation
\begin{equation}\label{sm12a}
    1-R-T-S=0.
\end{equation}

{\it - Scattering coefficients in the first-order Born approximation.}

The first-order Born approximation (FOBA) corresponds to the first term in the series expansion of $B(q)$ in the perturbation amplitude $\Delta\alpha$. FOBA can be simply obtained from Eq.(A4) by neglecting the integral term:
\begin{equation}\label{sm6}
    B(q)^{FOBA} =-\Delta\alpha(q-q_p).
\end{equation}
The FOBA is strictly valid only for small perturbations. However, it is fully analytical and, in many cases, it provides very useful information.
The calculation within FOBA for the defects considered in this manuscript are:

- Gaussian Defects. \\
In this case, a defect with width $a$ (defined as the full spatial width at $1/e$ relative conductivity change) has a conductivity profile $\Delta\alpha(x) = \delta \alpha_G \, \mathrm{exp}(-4x^2/a^2)$, where $\delta = (\alpha(x=0)-\alpha_G)/\alpha_G$. Then, $\Delta\alpha(q) = \frac{1}{4\sqrt{\pi}}\,  \alpha_G \,\tilde{a} \, \mathrm{exp}(-q^2 \tilde{a}^2/16)\, \delta $, with $\tilde{a} \equiv g a$.
According to Eq.(A8) and Eq.(A13) the reflectance reads
\begin{equation}\label{sm15}
R^{FOBA}=\frac{\pi}{4}  \, (k_p a)^2 \, e^{-\frac{1}{2} (k_p a)^2} \, \delta^2.
\end{equation}
where we have used that  $\alpha_G^{-2} = 1-q_p^2\approx -q_p^2$ .
The fraction of energy radiated out of plane can be estimated by approximating $B(q) \approx B(0)$ and
$G(q)\approx q_z$ for $|q|<1$. With this,
\begin{equation}\label{sm16}
S^{FOBA}\approx\frac{\pi}{8}  \, |\alpha_G|^2 \, (k_p a)^2 \, e^{-\frac{1}{8} (k_p a)^2} \, \delta^2.
\end{equation}
Notice that the reflectance maximum occurs at $k_p a = \sqrt{2}$. Then, at the reflectance maxima the FOBA predicts that the ratio $R/S= 2  e^{-3/4} \,  |\alpha_G|^{-2} \approx 1.8 \times 10^4 / (\alpha_G/\alpha_0)^2$,
where $\alpha_0 \approx 1/137$ is the fine structure constant.

- Abrupt defects. \\
For a defect with constant conductivity and spatial width $a$, $\Delta\alpha(x) = \delta \alpha_G \, \Theta(a/2-|x|)$. Then $\Delta\alpha(q) = \frac{1}{2 \pi}\, \alpha_G \, \tilde{a} \, \mathrm{sinc}(q \tilde{a}/2)\, \delta $, where
$\mathrm{sinc}(x) \equiv \mathrm{sin}(x)/x$. Using this expression, the scattering coefficients within the FOBA can be straightforwardly computed.

%
\providecommand*\mcitethebibliography{\thebibliography}
\csname @ifundefined\endcsname{endmcitethebibliography}
  {\let\endmcitethebibliography\endthebibliography}{}

\end{document}